\documentstyle[prd,aps]{revtex}
\newcommand{\sla}{\!\!\!/ \,}
\newcommand{\pa}{|{\bf p}|}
\newcommand{\kp}{{\bf k}\cdot {\bf p}}
\newcommand{\bga}{\mbox{\boldmath$\gamma$}}

\begin{document}
\draft

\title{Hard Loop Approach to Anisotropic Systems}

\author{Stanis\l aw Mr\' owczy\' nski\footnote{Electronic address:
mrow@fuw.edu.pl}}

\address{So\l tan Institute for Nuclear Studies\\
ul. Ho\.za 69, PL - 00-681 Warsaw, Poland\\
and Institute of Physics, Pedagogical University\\
ul. Konopnickiej 15, PL - 25-406 Kielce, Poland}

\author{Markus H. Thoma\footnote{Heisenberg fellow}\footnote{Electronic 
address: markus.thoma@cern.ch}}

\address{Theory Division, CERN \\
CH-1211 Geneva 23, Switzerland}

\date{17-th January 2000, revised 9-th February 2000}

\maketitle

\begin{abstract}

Anisotropic systems of quarks and gluons, which at least for 
sufficiently short space-time intervals can be treated as homogeneous 
and static, are considered. The gluon polarization tensor of such a
system is explicitly computed within the semiclassical kinetic and 
Hard Loop diagrammatic theories. The equivalence of  the two 
approaches is demonstrated. The quark self energy is computed as 
well, and finally, the dispersion relations of quarks and gluons in the 
anisotropic medium are discussed.

\end{abstract}


\section{Introduction}

The state of equilibrium being static and homogeneous is sometimes 
anisotropic. This may happen when the system of quantum fields, which
is of interest here, is under influence of an external force. A 
relativistic plasma, which is anisotropic due to a magnetic field, 
often occurs in astrophysical situations as the early Universe 
or Supernovae \cite{Zel83}. Anisotropic states are also common 
for systems which are out of equilibrium. Sometimes such states can be 
treated as static and homogeneous, but only for sufficiently short time 
and space intervals. How short the intervals should be depends on the 
specific problem under consideration. 

The parton system generated at the early stage of  ultrarelativistic heavy 
ion collisions at RHIC or LHC is of particular interest for us. The parton 
momentum distribution is not istotropic but strongly elongated along the 
beam \cite{Gei95,Wan97}. Therefore, specific color fluctuations, instead 
of being damped, can exponentially grow and noticeably influence the 
temporal evolution of the system. In a series of papers of one of us 
\cite{Mro93,Mro94} it has been argued that there are indeed very fast 
unstable plasma modes in such a parton system. The stability analysis 
\cite{Mro93,Mro94} has been performed within the semiclassical transport 
theory of quarks and gluons \cite{Elz89,Mro89}. Since the theory has been 
proven till now to be fully consistent with the QCD dynamics only for 
quasiequilibrium systems \cite{Bla94,Bla99} one wonders to what extend 
the results from \cite{Mro93,Mro94} are reliable. Thus, a QCD diagrammatic 
analysis is desirable. 

Perturbative approaches within the real time field theory provides 
a natural framework to study weakly interacting quantum field systems in and 
out of equilibrium. However, the naive perturbative expansion, when applied 
to gauge fields, suffers from various singularities and some physical 
quantities are even gauge dependent. These problems have been partly resolved 
for equilibrium systems by using an effective perturbative expansion where 
the so-called Hard Thermal Loops are resummed \cite{Bra90a}. The Hard Thermal 
Loop resummation technique within the finite-temperature QCD has been shown 
to be equivalent to the approach based on the classical \cite{Kel94} transport 
equations, where color is treated as a classical variable, or on the 
semiclassical \cite{Bla94} one, where the color degrees of freedom emerge due 
to the matrix structure of the parton distribution function. The Hard Thermal 
Loop approach has been generalized to nonequilibrium systems, but only very 
specific forms of deviations from the equilibrium have been discussed so far: 
systems out of chemical equilibrium, which are important in the context 
of heavy-ion collisions \cite{Bir93}, and such where the momentum 
distribution is isotropic but not of the Bose-Einstein or Fermi-Dirac form 
\cite{Bai97,LeB97,Car99}. As observed in \cite{Pis97}, the Hard Thermal Loop 
approach can be applied to any momentum distribution of hard particles which 
is static and homogeneous. This is evident when the Hard Thermal Loop 
effective action is derived within the transport theory \cite{Pis97}. 
The term `thermal' is then rather misleading and for this reason we shall 
omit it in the following. 

In this paper we discuss the applicability of the Hard Loop technique for 
systems with anisotropic momentum distributions. The technique has
been earlier applied to the equilibrium QED plasma in a magnetic field
\cite{Elm97}. Our aim is to consider a general situation with an arbitrary 
momentum distribution. We analyse the problem from the point of view 
of the transport theory and the diagrammatic approach. Using the 
semiclassical kinetic equations we derive the Hard Loop induced current 
paying much attention to the gauge aspects of the procedure. We also 
explicitly demonstrate that the gluon polarization tensors found by means 
of the two approaches are identical. In this way, the applicability of the 
kinetic theory beyond the equilibrium is substantiated and more specifically, 
the reliability of the results from \cite{Mro93,Mro94} is shown.

The Hard Loop diagrammatic technique has the advantage over the 
semiclassical transport theory approach that it can be naturally extended 
to fermionic self energies and to higher order diagrams beyond the 
semiclassical approximation. In this way the dispersion relations of quarks 
and other observables of the quark-gluon plasma, such as the energy loss 
of energetic partons, transport coefficients, or photon and dilepton 
production rates \cite{Tho95}, can be calculated systematically in the 
case of anisotropic distributions. We take a first step in this direction 
computing the quark self energy for an arbitrary momentum distribution. 

The self energy controls the particle dispersion relation which provides 
an essential dynamical information about the system. We discuss therefore 
the general dispersion relation of gluons (plasmons) and quarks in the 
anisotropic quark-gluon plasma. Finally, we briefly consider possible
applications of the formalism developed in this paper. 

\section{Transport Theory Approach}

In this section we first introduce the semiclassical transport theory of 
quarks and gluons \cite{Elz89,Mro89}. Then, applying the linear response 
method, the Hard Loop induced current is derived. Finally, we compute the 
gluon polarization. 

\subsection{Transport equations}

The distribution function of hard (anti-)quarks 
$Q({\bf p},x)\;\bigr(\bar Q({\bf p},x)\bigl)$ is a hermitian 
$N_c\times N_c$ matrix in color space (for a $SU(N_c)$ color 
group); $x$ denotes the space-time quark coordinate and ${\bf p}$ 
its momentum. The four-momentum $p=(E,{\bf p})$ is assumed to 
satisfy the mass-shell constraint. Since both quarks and gluons are 
treated as massless particles the constraint is $p^2 =0$. We also 
mention here that the spin of quarks and gluons is taken into 
account as an internal degree of freedom. The distribution function 
transforms under local gauge transformation $M$ as 
\begin{equation}\label{2.1a}
Q({\bf p},x) \rightarrow M(x)Q({\bf p},x)M^{\dag }(x) \;.
\end{equation}
The color indices are here and in the most cases below suppressed. The 
distribution function of hard gluons is a hermitian 
$(N_c^2-1)\times (N_c^2-1)$ matrix which transforms as
\begin{equation}\label{2.1b}
G({\bf p},x) \rightarrow {\cal M}(x)G({\bf p},x){\cal M}^{\dag }(x) \;,
\end{equation}
where
$$
{\cal M}_{ab}(x) = {\rm Tr}\bigr[\tau_a M(x) \tau_b M^{\dag }(x)]
$$
with $\tau_a ,\; a = 1,...,N_c^2-1$ being the $SU(N_c)$ group generators
in the fundamental representation. 

The color current is expressed in the fundamental representation as
\begin{eqnarray}\label{2.4}
j^{\mu }(x) = -g \int {d^3p \over (2\pi )^32E} \; p^{\mu } 
\Big[ Q({\bf p},x) - \bar Q ({\bf p},x) 
- {1 \over N_c}{\rm Tr}\big[Q({\bf p},x) 
&-& \bar Q ({\bf p},x)\big] \nonumber \\
&+&  2i \tau_a f_{abc} G_{bc}({\bf p},x)\Big] \;,
\end{eqnarray}
where $g$ is the QCD coupling constant, $f_{abc}$ are the structure 
constants of the $SU(N_c)$ group. 

The distribution functions of quarks and gluons are assumed to satisfy the 
following collisionless transport equations:
\begin{eqnarray}\label{2.5}
p^{\mu} D_{\mu}Q({\bf p},x) + g p^{\mu}{\partial \over \partial p_{\nu}}
{1 \over 2} \lbrace F_{\mu \nu}(x),Q({\bf p},x)\rbrace &=&  0\;, 
\nonumber \\
p^{\mu} D_{\mu}\bar Q({\bf p},x) - g p^{\mu}{\partial \over \partial p_{\nu}}
{1 \over 2} \{ F_{\mu \nu}(x),\bar Q({\bf p},x)\} &=& 0\;, 
\nonumber \\
p^{\mu} {\cal D}_{\mu}G({\bf p},x) + g p^{\mu}
{\partial \over \partial p_{\nu}}
{1 \over 2} \{ {\cal F}_{\mu \nu}(x),G({\bf p},x)\} &=& 0\;, 
\end{eqnarray}
where $\{...,...\}$ denotes the anicommutator; $D_{\mu}$ and 
${\cal D}_{\mu}$ are the covariant derivatives which act as
$$
D_{\mu} = \partial_{\mu} - ig[A_{\mu}(x),...\; ]\;,\;\;\;\;\;\;\;
{\cal D}_{\mu} = \partial_{\mu} - ig[{\cal A}_{\mu}(x),...\;]\;,
$$
with $A_{\mu }$ and ${\cal A}_{\mu }$ being the mean-field or background 
four-potentials;
$$
A^{\mu }(x) = A^{\mu }_a (x) \tau_a \;,\;\;\;\;\;
{\cal A}^{\mu }_{ab}(x) = - if_{abc}A^{\mu }_c (x) \;;
$$
$F_{\mu \nu}$ and ${\cal F}_{\mu \nu}$ are the mean-field stress
tensors with a color index structure analogous to that of the 
four-potentials. The background field is generated by the color current 
(\ref{2.4}) and the respective equation is
\begin{equation}\label{2.6}
D_{\mu} F^{\mu \nu}(x) = j^{\nu}(x)\;. 
\end{equation}
We note that the set of transport equations (\ref{2.5}, \ref{2.6}) is 
covariant with respect to the gauge transformations (\ref{2.1a}, \ref{2.1b}).

\subsection{Plasma color response}

We discuss here how the plasma, which is (on average) colorless,
homogeneous and stationary, responds to small color fluctuations.
The distribution functions are assumed to be of the form
\begin{eqnarray}\label{5.1}
Q_{ij}({\bf p},x) &=& n({\bf p})\delta_{ij} + \delta Q_{ij}({\bf p},x) \;, 
\\ \nonumber
\bar Q_{ij}({\bf p},x) &=& \bar n({\bf p})\delta_{ij} 
+ \delta \bar Q_{ij}({\bf p},x) \;,
\\ \nonumber
G_{ab}({\bf p},x) &=& n_g({\bf p})\delta_{ab} + \delta G_{ab}({\bf p},x) \;,
\end{eqnarray}
where the functions describing the deviation from the colorless state 
are assumed to be much smaller than the respective colorless functions.
The same is assumed for the momentum gradients of these functions.
The (anti-)quark and gluon distribution functions $ n({\bf p}) $, 
$\bar n({\bf p})$, $n_g({\bf p})$, reduce in equilibrium to the Fermi-Dirac 
or Bose-Einstein form i.e. 
\begin{eqnarray}\label{equilibrium}
n({\bf p}) &=& {2 \over \exp( \pa - \mu)/T+1} \;,
\\ \nonumber
\bar n({\bf p}) &=& {2 \over \exp( \pa + \mu)/T+1} \;,
\\ \nonumber
n_g({\bf p}) &=& {2 \over \exp (\pa/T) - 1} \;,
\end{eqnarray}
where $T$ and $\mu$ denote the temperature and quark chemical potential,
respectively, while the factor of 2 occurs due to the spin degrees of freedom.
The number of quark flavours is assumed to be equal to one.

Substituting (\ref{5.1}) in (\ref{2.4}) one gets
\begin{eqnarray}\label{5.2}
j^{\mu }(x) = -g \int {d^3p \over (2\pi )^32E}\, p^{\mu} 
\Bigr[ \delta Q({\bf p},x) - \delta \bar Q ({\bf p},x) 
-{1 \over N_c}Tr\bigr[\delta Q({\bf p},x) 
&-& \delta \bar Q ({\bf p},x)\bigl]
\\ \nonumber
&+& 2i \tau_a f_{abc} \delta G_{bc}({\bf p},x)\Bigl] \;.
\end{eqnarray}
As seen, the current occurs due to the deviation from the colorless 
state. Let us also observe here that not only (anti-)quarks but also 
gluons contribute to the current (\ref{5.2}). Thus, the current
is of essentially non-Abelian nature.

Now, we substitute the distribution functions (\ref{5.1}) into the 
transport equations (\ref{2.5}). Assuming that the stress tensor is 
of the same order as  $\delta Q$, $\delta \bar Q$ or $\delta G$ and
linearizing the equations with respect to $\delta Q$, $\delta \bar Q$ 
and $\delta G$ we get
\begin{eqnarray}\label{5.4}
p^{\mu} D_{\mu} \delta Q({\bf p},x) 
&=& - g p^{\mu}F_{\mu \nu}(x){\partial n({\bf p}) \over \partial p_{\nu}}\;, 
\\ \nonumber
p^{\mu} D_{\mu} \delta \bar Q({\bf p},x) 
&=& \;\;\;
g p^{\mu}F_{\mu \nu}(x){\partial \bar n({\bf p}) \over\partial p_{\nu}}\;, 
\\ \nonumber
p^{\mu} {\cal D}_{\mu} \delta G({\bf p},x) 
&=&- 
g p^{\mu}{\cal F}_{\mu \nu}(x){\partial n_g({\bf p}) \over \partial p_{\nu}}\;.
\end{eqnarray}
We keep here the covariant derivatives to maintain the gauge covariance of
the equations.

To solve the equations such as Eqs. (\ref{5.4}) one usually uses, see e.g.
\cite{Elz89,Bla94}, the gauge parallel transporter defined in the 
fundamental representation as
$$
U(x,y) = {\cal P}\: {\rm exp}
\Big[ -ig \int_x^y dz_{\mu} \: A^{\mu}(z) \Big] \;,
$$
where ${\cal P}$ denotes the ordering along the path from $x$ to $y$. 
There is analogous formula of the gauge transporter ${\cal U}(x,y)$ in
the adjoint representation. Using $U$ and ${\cal U}$ one finds the solutions 
of Eqs. (\ref{5.4}) as
\begin{eqnarray}\label{5.5}
\delta Q({\bf p},x) &=& - g \int d^4 y \:
G_p(x - y) \; U(x,y)\;
p^{\mu}F_{\mu \nu}(y) \; 
U(y,x)\;{\partial n({\bf p}) \over \partial p_{\nu}} \;, 
\\ \nonumber
\delta \bar Q({\bf p},x) &=& \;\;\; g \int d^4 y \:
G_p(x - y) \; U(x,y)\; 
p^{\mu}F_{\mu \nu}(y)\;
U(y,x)\;{\partial \bar n({\bf p}) \over \partial p_{\nu}}\;, 
\\ \nonumber
\delta G({\bf p},x) &=& - g \int d^4 y \:
G_p(x - y) \; {\cal U}(x,y)\;
p^{\mu}{\cal F}_{\mu \nu}(y) \; {\cal U}(y,x)
\;{\partial n_g({\bf p}) \over \partial p_{\nu}} \;, 
\end{eqnarray}
where $G_p(x)$ is the retarded Green's function which satisfies the 
equation 
$$
p_{\mu}\partial^{\mu} G_p(x) = \delta^{(4)}(x)
$$
and equals
$$
G_p(x) = E^{-1} \Theta (t) \,  \delta^{(3)} ({\bf x} - {\bf v}t) \;,
$$
with $t$ being the 0-th component of $x$ ($x^{\mu} \equiv (t, {\bf x})$),
and ${\bf v}$ denoting the parton velocity i.e. ${\bf v} \equiv  {\bf p}/E $.

Substituting the solutions (\ref{5.5}) in Eq. (\ref{5.2}) one finds the color 
current of the gauge covariant form which reads
\begin{eqnarray}\label{ind-current}
j^{\mu }(x) =  g^2 \int {d^3p \over (2\pi )^32E}\, p^{\mu}p^{\lambda} 
\int d^4y \: G_p(x - y) \; U(x,y)\; 
F_{\lambda \nu}(y) \; U(y,x)\;{\partial f({\bf p})\over \partial p_{\nu}}
\end{eqnarray}
where $f({\bf p}) \equiv n({\bf p}) + \bar n({\bf p})  + 2N_c n_g({\bf p})$. 

Now, we are going to perform the Fourier transform of the induced current 
(\ref{ind-current}). Before this step however, we neglect the terms which are 
not of the leading order in $g$. Then, the transporters $U$ are approximated 
by unity and the stress tensor $F_{\mu \nu}$ by 
$\partial_{\mu} A_{\nu} - \partial_{\nu} A_{\mu} $. Within such an 
approximation, the Fourier transformed induced current (\ref{ind-current}),
which is no longer gauge covariant, equals 
\begin{equation}\label{ind-current-k}
j^{\mu}(k) = g^2  \int {d^3p \over (2\pi )^3 2E} \: p^{\mu}
{\partial f ({\bf p}) \over \partial p_{\lambda}} \Bigg[ g^{\lambda \nu} 
- { k^{\lambda} p^{\nu} \over p^{\sigma}k_{\sigma} + i0^+} \Bigg] \;
A_{\nu}(k) \;.
\end{equation}

The induced current $ j^{\mu}(k) $ can be expressed as
$$
j^{\mu}_a(k) = \Pi^{\mu \nu}_{ab}(k) A_{\nu}^{b}(k) \;,
$$
with $\Pi^{\mu \nu}$ being the gluon polarization tensor. 
Transforming Eq.~(\ref{ind-current-k}) to the adjoint 
representation one finds
\begin{equation}\label{Pi}
\Pi^{\mu \nu}(k) = g^2  \int {d^3p \over (2\pi )^3 2E} \: p^{\mu}
{\partial f ({\bf p}) \over \partial p_{\lambda}} \Bigg[ g^{\lambda \nu} 
- { k^{\lambda} p^{\nu} \over p^{\sigma}k_{\sigma} + i0^+} \Bigg] \;.
\end{equation}
It should be noted here that the polarization tensor is proportional to 
a unit matrix in the color space.

Now we are going to show that the polarization tensor is transversal i.e.
$k_{\mu} \Pi^{\mu \nu}(k) = 0$. Let us first consider 
$ k_{\mu} \Pi^{\mu 0}(k) $. One immediately finds from Eq.~(\ref{Pi}) that
$$
k_{\mu} \Pi^{\mu 0}(k) = - {g^2 \over 2} \: k^l \int {d^3p \over (2\pi )^3}
{\partial f ({\bf p}) \over \partial p^l}  \;.
$$
The indices $l,m,n = 1,2,3$ refer to the coordinates of three-vectors.
The energy density carried by partons is expected to be finite. Therefore,
$f({\bf p}= \infty)$ must vanish. Consequently, the above integral vanishes as
well. Performing partial integration and demanding that $f({\bf p}= \infty)=0$ 
one also proves that $k_{\mu} \Pi^{\mu m}(k) = 0$. Analogously it can be
also shown that $\Pi^{\mu \nu}(k) = \Pi^{\nu \mu}(k)$.

\section{Diagrammatic approach}

In this section we consider the diagrammatic Hard Loop approach to 
anisotropic systems. Specifically, we compute the QCD polarization tensor 
and the quark self energy for an arbitrary parton momentum distribution.

\subsection{Polarization tensor}

The contribution from the quark loop to the gluon self energy in the case 
of one quark flavor is of the form
\begin{equation} \label{pol1}
\Pi^{\mu \nu}_{ab} (k) = \frac{i}{2}\> \delta_{ab} \> 
g^2 \> \int \frac{d^4p}{(2\pi )^4}\> 
{\rm Tr} [\gamma ^\mu S(q)\gamma ^\nu S(p)]\; ,
\end{equation}
where $q \equiv p-k$ and $S$ is the bare quark propagator. 
Since we are dealing with a non-equilibrium situation we adopt the 
real time formalism. Within the Keldysh representation \cite{Cho85},
which has been shown to be especially convenient in the Hard Loop 
approximation \cite{Car99}, there are retarded $(R)$, advanced $(A)$, and 
symmetric $(F)$ propagators which in the case of massless quarks are given by
\begin{eqnarray} \label{prop}
S_{R,A}(p) & = & \frac{p\sla}{p^2\pm i\, {\rm sgn}(p_0)0^+}\nonumber \; ,\\
S_F(p) & = & -2\pi i\> p\sla \Big( [1- n({\bf p})] \Theta(p_0)
+ [1- \bar n({\bf p})] \Theta(- p_0) \Big) \delta (p^2)\; ,
\end{eqnarray}
where $n({\bf p})$ ($\bar n({\bf p})$)  is, as previously, the (anti-)quark 
distribution function \cite{LeB97} that reduces in equilibrium to  the form 
(\ref{equilibrium}). Performing the trace in (\ref{pol1}) and suppressing 
the color indices, we find the retarded gluon self energy as
\begin{equation} \label{pol2}
\Pi ^{\mu \nu} (k) = i\> g^2 \> \int \frac{d^4p}{(2\pi )^4}\> 
[q^\mu p^\nu + p^\mu q^\nu - g^{\mu \nu} (q\cdot p)]\>[\tilde \Delta_F(q)
\tilde \Delta_R(p) + \tilde \Delta_A(q) \tilde \Delta_F(p)]\; ,
\end{equation}
where $S_{R,A,F}(p)=p\sla \tilde \Delta_{R,A,F}(p)$. Terms containing
$\tilde \Delta_A(q)\tilde \Delta_A(p)$ and 
$\tilde \Delta_R(q) \tilde \Delta_R(p)$ have been neglected as they 
vanish after integrating over $p_0$.

First, we will consider the spatial components of the polarization tensor.
The other components follow from it using the transversality of the
polarization tensor, as we will discuss below. After performing the 
integration over $p_0$, we obtain
\begin{eqnarray} \label{pol3}
\Pi^{lm}(k)=- {g^2 \over 2} \> \int \frac{d^3p}{(2\pi )^3}\>
\frac{f_q({\bf p})}{\pa}\> && \Biggl [\frac{2p^lp^m-k^lp^m-p^lk^m+
\delta^{lm}(-\omega \pa +\kp)}{-2\omega \pa +2\kp +k^2
-i\, {\rm sgn}(\pa-\omega)0^+}\nonumber \\
&& + \frac{2p^lp^m-k^lp^m-p^lk^m+
\delta^{lm}(\omega \pa +\kp)}{2\omega \pa +2\kp +k^2
-i\, {\rm sgn}(-\pa-\omega)0^+}\Biggr ]\; ,
\end{eqnarray}
where $\omega$ is the 0-th component of $k$ i.e. $k=(\omega, {\bf k})$
and $f_q({\bf p}) \equiv n({\bf p})+\bar n({\bf p})$. Here the vacuum part
has been neglected because it is suppressed compared to the matter part in 
the Hard Loop approximation.

Adopting the Hard Loop approximation we assume that the internal momenta 
are much larger than the external one, i.e. $\omega $, $k_l\ll p_l $. Note 
that for arbitrary anisotropic distributions we have to require the Hard Loop
condition for each component of the momenta, whereas in the isotropic case
$\omega $, $|{\bf k}|\ll \pa$ suffices. Expanding the expression in the square 
brackets for small external momenta yields
\begin{eqnarray} \label{expand}
&& \frac{p^lp^m}{-\omega \pa +\kp-i0^+} 
+\frac{p^lp^m}{\omega \pa +\kp+i0^+}
+\frac{-k^lp^m-p^lk^m+\delta^{lm}(-\omega \pa +\kp)}{2(-\omega \pa +\kp-i0^+)}
\nonumber \\
&&
+\frac{-k^lp^m-p^lk^m+\delta^{lm}(\omega \pa +\kp)}{2(\omega \pa +\kp+i0^+)}  
-\frac{p^lp^m\> k^2}{2(-\omega \pa +\kp-i0^+)^2}
-\frac{p^lp^m\> k^2}{2(\omega \pa +\kp+i0^+)^2}.
\end{eqnarray}
In equilibrium the first two terms vanish after integrating over
${\bf p}$. This also holds out of equilibrium if we assume 
$f_q(-{\bf p})=f_q({\bf p})$. Then, we arrive at the final result
\begin{equation} \label{pol4}
\Pi^{lm}(k)=-{g^2 \over 2} \> \int \frac{d^3p}{(2\pi )^3}\>
\frac{f({\bf p})}{\pa}\> \frac{(k^lp^m+p^lk^m)\,
(\omega \pa -\kp)+\delta^{lm}(\omega \pa -\kp)^2
-p^lp^m(\omega^2-|{\bf k}|^2)}{(\omega \pa -\kp+i0^+)^2}\; ,
\end{equation}
where we replaced $f_q({\bf p})$ by 
$f({\bf p}) \equiv n({\bf p})+ \bar n({\bf p}) +2N_c n_g({\bf p})$.
The point is that in the Hard Loop limit the gluonic contributions to 
the polarization tensor have the same structure as the quark ones 
\cite{Pes98} and only the distribution function and the color factor
change. For essentially the same reason, the QCD polarization tensor,
computed even with the complete expression (\ref{expand})
without assuming $f(-{\bf p}) = f({\bf p})$,  is gauge 
independent in the Hard Loop approximation. Indeed, the gluon 
polarization tensor has the same structure as the photon one in the 
Hard Loop limit. Since the one-loop photon polarization tensor 
contains no gauge boson propagator it is gauge independent. 
Consequently, the same holds for the gluon polarization tensor.
The result (\ref{pol4}) is fully equivalent to Eq.~(\ref{Pi}) obtained 
within the semiclassical kinetic theory. In order to show the equivalence, 
one performs a partial integration in (\ref{Pi}) and immediately gets 
Eq.~(\ref{pol4}). However, we do not need to assume the reflection 
symmetry of the distribution function to derive Eq.~(\ref{Pi}). 

Two more comments are in order here. First, if we do not assume the 
reflection symmetry of  $f({\bf p})$ the first two terms in (\ref{expand}) 
will contribute, leading to contributions in the polarisation tensor that 
dominate over the Hard Loop result (\ref{pol4}) and are absent in 
the semiclassical approximation. As explained above, the extra 
contribution appears to be gauge independent. Second, for equilibrium 
distribution functions (\ref{equilibrium}) the integrals over $\pa $ and over 
the angle in (\ref{pol4}) factorize. Then, it is easy to show that (\ref{pol4}) 
reduces to the well known Hard Thermal Loop result \cite{Kli82} where 
the polarization tensor has only two independent components and depends 
on $\omega $ and $|{\bf k}|$.

Owing to transversality, the time-like components of $\Pi^{\mu \nu}$
follow from $\Pi^{lm}$. Indeed, $\Pi^{0m}(k)=k^l\Pi^{lm}(k)/\omega$ and 
$\Pi^{00}(k)=k^lk^m\Pi^{lm}(k)/\omega ^2$. In order to prove the 
transversality of the hard loop polarization tensor in the case of anisotropic 
distributions, we compute $k_\mu \Pi^{\mu \nu}(k)$. Considering first the 
quark loop contribution, we get 
\begin{equation} \label{trans1}
k_\mu \Pi ^{\mu \nu} (k) = i\> g^2 \> \int \frac{d^4p}{(2\pi )^4}\> 
\big[ 2 (k\cdot p) p^\nu - p^2 k^\nu - k^2 p^\nu \big] \> 
\big[ \tilde \Delta_F(q) \tilde \Delta_R(p) 
+ \tilde \Delta_A(q) \tilde \Delta_F(p)\big]\; .
\end{equation}
After integrating over $p_0$ we find
\begin{eqnarray} \label{trans2}
k_\mu \Pi^{\mu 0}(k)=- {g^2 \over 2} \> \int 
\frac{d^3p}{(2\pi )^3}\> \frac{f_q({\bf p})}{\pa}\> 
&& \Biggl [\frac{2(\omega \pa -\kp )\pa -k^2\pa}
{-2\omega \pa +2\kp +k^2-i\, {\rm sgn}(\pa -\omega)0^+}\nonumber \\
&& + \frac{-2(-\omega \pa - \kp)\pa +k^2\pa}
{2\omega \pa +2\kp +k^2-i\, {\rm sgn}(-\pa -\omega)0^+}\Biggr ]
\end{eqnarray}
and
\begin{eqnarray} \label{trans3}
k_\mu \Pi^{\mu m}(k)=- {g^2 \over 2} \> \int 
\frac{d^3p}{(2\pi )^3}\> \frac{f_q({\bf p})}{\pa}\> 
&& \Biggl [\frac{2(\omega \pa -\kp )p^m-k^2p^m}
{-2\omega \pa +2\kp +k^2-i\, {\rm sgn}(\pa -\omega)0^+}\nonumber \\
&& + \frac{2(-\omega \pa - \kp)p^m -k^2p^m}
{2\omega \pa +2\kp +k^2-i\, {\rm sgn}(-\pa -\omega)0^+}\Biggr ]\; .
\end{eqnarray}
Expanding the integrands in the these expressions for small external momenta 
analogously to (\ref{expand}), it is easy to show that (\ref{trans2})
and (\ref{trans3}) vanish in the Hard Loop approximation. This also holds 
if the gluon loop contribution is added as they have the same structure 
in the Hard Loop approximation.

\subsection{Fermion self energy}

As mentioned in the Introduction, the diagrammatic technique has the advantage 
over the semiclassical transport theory approach that it can be easily 
extended to fermionic self energies. Therefore, we discuss the Hard Loop 
quark self energy for anisotropic momentum distributions. Using the Feynman 
gauge, the one-loop quark self energy is found as
\begin{equation} \label{self1}
\Sigma_{ij}(k) = 2i \, C_F \> \delta_{ij}\>
g^2 \> \int \frac{d^4p}{(2\pi )^4}\> 
S(p)\> \Delta(q)\; ,
\end{equation}
where $C_F \equiv (N_c^2 - 1)/N_c$ and now $q \equiv k-p$. Adopting again the 
Keldysh representation, the gluon propagators in the Feynman gauge are given by
\begin{eqnarray} \label{prop1}
\Delta_{R,A}(q) & = & \frac{1}{q^2\pm i {\rm sgn}(q_0)0^+}\nonumber \; ,\\
\Delta_F(q) & = & -2\pi i\> \big[1+n_g ({\bf q}) \big] \;\delta (q^2)\; .
\end{eqnarray}
Suppressing the color indces, we find for the retarded quark self energy as
\begin{equation} \label{self2}
\Sigma (k) = i\, g^2 \, C_F \int \frac{d^4p}{(2\pi )^4}\> 
p\sla \> \Big[\tilde \Delta_R(p)\Delta_F(q) + \tilde \Delta_R(p)\Delta_A(q)
+ \tilde \Delta_F(p)\Delta_R(q) + \tilde \Delta_A(p)\Delta_R(q)\Big]\; .
\end{equation}
The matter part of (\ref{self2}) can be decomposed in two contributions
which read after integrating over $p_0$
\begin{eqnarray} \label{self3}
\Sigma_1(k) = \frac{g^2}{8} \, C_F
 \int \frac{d^3p}{(2\pi )^3}\>
\frac{n_g({\bf p})}{\pa}\> &&\Biggl [\frac{(\omega -p)\gamma_0-
({\bf k}-{\bf p})\cdot \bga}{-2\omega \pa +2\kp +k^2
+i{\rm sgn}(\omega -\pa)0^+}\\ \nonumber 
&&+ \frac{(\omega +p)\gamma_0-
({\bf k}-{\bf p})\cdot \bga}{2\omega \pa +2\kp +k^2
+i{\rm sgn}(\omega +\pa)0^+}\Biggr]
\end{eqnarray}
and
\begin{eqnarray}
\Sigma_2(k) = \frac{g^2}{16} \, C_F
 \int \frac{d^3p}{(2\pi )^3}\>
\frac{ n({\bf p})  + \bar n({\bf p})  }{\pa}\> 
&&\Biggl[ \frac{-p\gamma_0+
{\bf p}\cdot \bga}{-2\omega \pa +2\kp +k^2
+i{\rm sgn}(\omega -\pa)0^+}\\ \nonumber 
&&+ \frac{p\gamma_0+
{\bf p}\cdot \bga}{2\omega \pa +2\kp +k^2
+i{\rm sgn}(\omega +\pa)0^+}\Biggr ]\; .
\end{eqnarray}
In contrast to the polarization tensor, we need to expand the square brackets 
in (\ref{self3}) only to the first order for small external momenta, leading to
\begin{equation}\label{expand1}
\frac{-p\gamma_0+ {\bf p}\cdot \bga }{-2\omega \pa +2\kp-i0^+} 
+ \frac{p\gamma_0+{\bf p}\cdot \bga}{2\omega \pa +2\kp+i0^+}\; .
\end{equation}
Assuming again the reflection symmetry for the distribution functions, 
we obtain the final gauge independent result in the Hard Loop approximation 
\begin{equation} \label{self4}
\Sigma (k) = \frac{g^2}{16}C_F \int \frac{d^3p}{(2\pi )^3}\>
\frac{2 n_g({\bf p})+ n({\bf p})  + \bar n({\bf p}) }{\pa }\> 
\frac{\gamma _0+{\bf v}\cdot \bga}
{\omega +{\bf v}\cdot {\bf k}+i0^+}\; .
\end{equation}
In the case of isotropic distributions (\ref{self4}) reduces to the well known 
Hard Thermal Loop result \cite{Wel82}, where the self energy for massless 
quarks contains only two independent scalar functions depending on $\omega$ 
and $|{\bf k}|$. Giving up the reflection symmetry of the distribution 
functions does not introduces new dominant terms in this case since the self 
energy follows already from the lowest order terms (\ref{expand1}).
We have adopted this symmetry to treat $\Pi$ and $\Sigma$ in exactly
the same way.

\section{Dispersion relations}

The gluon polarization tensor and quark self energy can be used to determine
the dispersion relations of gluons (plasmons) and quarks in the quasistatic
and quasihomogeneous but anisotropic state of the quark-gluons plasma.

\subsection{Gluon dispersion equation}

The background gluon field $A^{\mu}(k)$ satisfies the following equation 
of motion
$$
\Big[ k^2 g^{\mu \nu} -k^{\mu} k^{\nu} - \Pi^{\mu \nu}(k) \Big] 
A_{\nu}(k) = 0 \;.
$$
Therefore, the general plasmon dispersion equation is of the form 
\begin{equation}\label{dispersion-pi}
{\rm det}\Big[ k^2 g^{\mu \nu} -k^{\mu} k^{\nu} - \Pi^{\mu \nu}(k) \Big] 
 = 0 \;.
\end{equation}
Equivalently, the dispersion relations are given by the positions of the
pole of  the effective gluon propagator. Due to the transversality of 
$\Pi^{\mu \nu}$ not all components of $\Pi^{\mu \nu}$ are independent 
from each other and consequently the dispersion equation (\ref{dispersion-pi}) 
can be simplified. For this purpose we introduce the color permittivity tensor 
$\epsilon^{lm}(k)$. Because of the relation 
$$
\epsilon^{lm}(k) E^l(k) E^m(k) = \Pi^{\mu \nu} (k) A_{\mu}(k) A_{\nu}(k)\;,
$$
where ${\bf E}$ is the chromoelectric vector, the permittivity can be 
expressed through the polarization tensor as
\begin{equation} \label{epspol}
\epsilon^{lm}(k) = \delta^{lm} + {1 \over \omega^2} \Pi^{lm}(k) \;.
\end{equation}
There are two other equalities which follow from the transversality of 
$\Pi^{\mu \nu}$. Namely,
$$
\Pi^{00}(k)  = (\epsilon^{lm}(k) - \delta^{lm})\: k^l k^m \;,
\;\;\;\;\;\;\;\;\;\;\;\;\;\;\;\;\;\;
\Pi^{l0}(k) = (\epsilon^{lm}(k) - \delta^{lm}) \: \omega \: k^m \;.
$$
Using the permittivity tensor the dispersion equation gets the form
\begin{equation}\label{dispersion-g}
{\rm det}\Big[ {\bf k}^2 \delta^{lm} -k^l  k^m 
- \omega^2 \epsilon^{lm}(k)  \Big]  = 0 
\end{equation}
with 
\begin{equation}\label{epsilon}
\epsilon^{lm}(k) = \delta^{lm} + {g^2 \over 2 \omega} 
\int {d^3p \over (2\pi )^3} 
\:{ v^l  \over \omega - {\bf k \cdot v} + i0^+}
{\partial f ({\bf p}) \over \partial p^n} \;
\bigg[ \Big(1 - {{\bf k \cdot v} \over \omega} \Big) \delta^{nm}
+ {k^n v^m \over \omega} \bigg] \;.
\end{equation}

In the isotropic state there are only two independent components of the
permittivity tensor
$$
\epsilon^{lm}(k) = \epsilon_T (k) \Big(\delta^{lm} - k^l k^m/{\bf k}^2)\Big)
+ \epsilon_L (k) \; k^l k^m/{\bf k}^2 \;,
$$
and the dispersion equation (\ref{dispersion-g}) splits into two equations
$$
\epsilon_T (k) = {\bf k}^2/ \omega^2 \;,\;\;\;\;\;\;\;\;\;\;\;
\epsilon_L (k) = 0 \;.
$$

The permittivity tensor (\ref{epsilon}) was calculated for the strongly 
elongated parton momentum distribution $f({\bf p})$ and it was found 
\cite{Mro93,Mro94} that there are unstable solutions of the dispersion 
equation (\ref{dispersion-g}).

\subsection{Quark dispersion equation}

The quark dispersion relations are determined by the poles of the 
hard loop resummed quark propagator or equivalently are found 
as solutions of the equation
\begin{equation}\label{dispersion-q1}
{\rm det}\Big[ p\sla  - \Sigma (p)  \Big]  = 0 \;.
\end{equation}
One sees in Eq.~(\ref{self4}) that the spinor structure of $\Sigma$ 
is very simple: $\Sigma (p) = \gamma^{\mu} \Sigma_{\mu}(p)$.
However, we also include here the scalar part which is relevant for 
the massive quarks. Then, 
\begin{equation}\label{structure1}
\Sigma (p) = \gamma^{\mu} \Sigma_{\mu}(p)  +  C(p) \;.
\end{equation}
Substituting the expression (\ref{structure1}) into 
Eq.~(\ref{dispersion-q1}) and computing the determinant as 
explained in Appendix 1 of \cite{Mro94b}, we get
\begin{equation}\label{dispersion-q2}
\Big[\big( p^{\mu} - \Sigma^{\mu}(p) \big) 
\big(p_{\mu} - \Sigma_{\mu}(p) \big)  - C^2(p) \Big]^2  = 0 \;.
\end{equation}

When the momentum distribution is isotropic, the structure of 
$\Sigma$ further simplifies \cite{Wel82}:
$$
\Sigma (p) = A(p) \, p_0 \gamma^0 + B(p)\, {\bf p}\cdot \bga 
+ C(p) \;.
$$
Then, the dispersion equation reads
$$
\big(1 - A(p) \big)^2 p_0^2 - \big(1 - B(p) \big)^2 {\bf p}^2 
- C^2(p) = 0 \;.
$$

\section{Discussion}

In the present work we have considered an anisotropic relativistic plasma 
which is either in equilibrium and the anisotropy is caused by external fields
or the plasma is out of equilibrium. In the first case we deal with the 
homogeneous and static systems while in the second one it can be treated as
quasihomogenous and quasistatic for sufficiently short space-time intervals. 
An example of the first case is the magnetized plasma while of the second 
one the parton system from the early stage of relativistic heavy-ion 
collisions where we encounter a strong anisotropy in the momentum 
distribution. 

The QCD polarization tensor has been computed in two ways. We have first
applied the linear response method within the semiclassical transport theory 
and then the diagrammatic Hard Loop approach. The two methods are equivalent 
(but the distribution functions have to possess a reflection symmetry, i.e. 
$f(-{\bf p})=f({\bf p})$). When using the diagrammatic approach
we have referred to the real time formalism since the systems under
consideration are, in general, out of equilibrium. According to the Hard 
Loop approximation, we have used bare propagators for the internal lines 
of  the polarization tensor which exhibit an explicit anisotropic momentum 
distribution. Another method has been used in \cite{Elm97,Ken98} to study 
anisotropic relativistic QED plasmas in a strong magnetic field. The system 
has been assumed there to be in equilibrium and dressed propagators 
corresponding to electrons in Landau levels have been adopted. In this way, 
anisotropic distributions arise although the distribution functions depend 
only on the energy. 

As already mentioned, the semiclassical kinetic theory of quarks and gluons
has been shown so far to be fully consistent with the QCD dynamics only for 
quasiequilibrium \cite{Bla94,Bla99}. The considerations presented here 
demonstrate that the equivalence holds for the systems which are far from 
equilibrium although the space-time homogeneity must be invoked. Thus, the 
reliability of the kinetic theory methods is improved and in particular, 
the stability analysis of the parton system form the early stage of 
ultrarelativistic heavy-ion collisions, which has been based on the linearized 
kinetic equations, \cite{Mro93,Mro94} is substantiated. 

The main advantage of the diagrammatic approach over the transport one 
is that it allows for a systematic perturbative extension to higher order 
effects. Also the fermion self energy for anisotropic systems can be 
calculated in this way.  Having the QCD polarization tensor and quark self 
energy derived here, one can construct effective gluon and quark propagators 
from the Dyson-Schwinger equation. 

The poles of the effective propagators determine 
(via Eqs.~(\ref{dispersion-g}, \ref{dispersion-q2})) the parton dispersion 
relations in an anisotropic quark-gluon plasma. In the isotropic plasma, the 
dispersion relations for gluons and quarks show two branches and start from 
the same energy at zero momentum \cite{Kli82}. The point is that in this case
there is no direction preferred and the longitudinal and transverse components 
of the dielectric function are identical when the momentum vanishes. For the 
anisotropic systems with a preferred direction even at zero momentum, we 
expect additional branches and the degeneracy at zero momentum to be 
removed. In equilibrium all modes are stable or damped due to the 
Landau mechanism. In the case of anisotropic systems growing modes, i.e.  
instabilities,  are possible. The unstable modes were argued to occur in the 
parton system from the early stage of ultrarelativistic heavy-ion collisions 
\cite{Mro93,Mro94}. Since the characteristic time of instability development 
was estimated to be rather small (below 1 fm/c) these instabilities can 
significantly influence the temporal evolution of the parton system.

Also the quark dispersion relations following from the effective quark
propagator are of physical relevance, as they lead in equilibrium to
interesting structures, e.g. van Hove peaks, in the dilepton production 
rate \cite{Bra90b}, which might serve as a signature for the quark-gluon 
plasma formation. It has to be seen whether these structures also survive 
in the nonequilibrium case.

When the plasma is in the (isotropic) equilibrium state the zero frequency 
limit of the longitudinal component of the polarization tensor in the Hard 
Thermal Loop limit ($\Pi_L=\Pi_{00}$), which is identified with the lowest 
order Debye screening mass, is finite. The transverse component 
($\Pi_T=(\delta_{lm}-k_lk_m/|{\bf k}|^2)\Pi_{lm}/ 2$), on the other hand,
shows no static magnetic screening. The situation is much more complicated
in the anisotropic plasma. The screening length depends on the orientation
of the vector ${\bf k}$ \cite{Mro94}, see also \cite{Bir92}. 

The diagrammatic approach, following the Hard Loop resummation technique 
\cite{Bra90a,Car99}, allows for a systematic calculation of observables, 
such as the energy loss of energetic partons or the production of photons 
and dileptons \cite{Tho95}. Now, the program can be extended to the 
anisotropic quark-gluon plasma although one has to choose a specific form 
of the parton momentum distribution. We leave all these issues for future 
investigations. Here we have intended to provide only the general formalism 
to study anisotropic systems of quantum fields.

\begin{acknowledgments}

We are very grateful to the National Institute of Nuclear Theory at Seattle
where this project was initiated during the Workshop Non-equilibrium 
Dynamics in Quantum Field Theory. 

\end{acknowledgments}  


\end{document}